\documentclass[sigconf,  review=false, authorversion]{acmart}
\usepackage{graphicx}
\usepackage[utf8]{inputenc} 
\usepackage{comment}
\usepackage{glossaries}
\usepackage{mathtools}
\usepackage[ruled,vlined]{algorithm2e}
\usepackage{tikz}
\usepackage{tikz-uml}
\usetikzlibrary{shapes,arrows}
\usetikzlibrary{positioning}
\usetikzlibrary{fit}
\usetikzlibrary{decorations.pathreplacing}

\usepackage[htt]{hyphenat} 

\usepackage{cleveref}
\usepackage{siunitx}

\newglossaryentry{param}
{
  name={\ensuremath{\theta}},
  description={Parameter vector}
}
\newglossaryentry{paramdim}
{
  name={\ensuremath{m}},
  description={Parameter dimension}
}
\newglossaryentry{numsamps}
{
  name={\ensuremath{N}},
  description={Number of samples}
}
\newglossaryentry{propdist}
{
  name={\ensuremath{q}},
  description={Proposal distribution}
}
\newglossaryentry{priordist}{
name=\ensuremath{\pi},
description={Prior distribution},
}
\newglossaryentry{qoi}{
name=\ensuremath{Q},
description={Quantity of interest},
}
\newglossaryentry{meas}{
name=\ensuremath{y},
description={Measurement},
}
\newglossaryentry{posteriordist}{
name=\ensuremath{\mathcal{P}(\gls{param}|\gls{meas})},
description={Measurement distribution},
}
\newglossaryentry{posteriordens}{
name=\ensuremath{\nu},
description={Measurement distribution},
}
\newglossaryentry{likelihood}{
name=\ensuremath{\mathcal{L}},
description={Likelihood function},
}
\newglossaryentry{measdist}{
name=\ensuremath{\mathcal{P}(y)},
description={Measurement distribution},
}
\newglossaryentry{cost}{
name=\ensuremath{\mathcal{C}},
description={Measurement distribution},
}

\newacronym{UQ}{UQ}{Uncertainty Quantification}
\newacronym{PDE}{PDE}{partial differential equations}
\newacronym{MC}{MC}{Monte Carlo}
\newacronym{MAP}{MAP}{maximum a posteriori probability}
\newacronym{MCMC}{MCMC}{Markov Chain Monte Carlo}
\newacronym{MHMCMC}{MHMCMC}{Metropolis-Hastings Markov Chain Monte Carlo}
\newacronym{MLMCMC}{MLMCMC}{Multilevel Markov Chain Monte Carlo}
\newacronym{MLMC}{MLMC}{Multilevel Monte Carlo}
\newacronym{HPC}{HPC}{High Performance Computing}
\newacronym{ADER-DG}{ADER-DG}{Arbitrary-high-order-DERivative Discontinuous Galerkin Method}
\newacronym{ExaHyPE}{ExaHyPE}{Exascale Hyperbolic PDE Engine}
\newacronym{MUQ}{MUQ}{MIT Uncertainty Quantification Library}
\newacronym{ODE}{ODE}{ordinary differential equations}
\newacronym{QOI}{QOI}{Quantity of Interest}
\newacronym{MPI}{MPI}{Message Passing Interface}
\newacronym{TBB}{TBB}{Intel Threading Building Blocks}
\newacronym{DUNE}{DUNE}{Distributed and Unified Numerics Environment}
\newacronym{KL}{KL}{Karhunen Lo\`{e}ve}

\usepackage{spverbatim}
\usepackage{listings}
\begin{document}

 \setcopyright{none}
 \copyrightyear{2021}
 \acmYear{2021}
 \acmDOI{TODO}
%
 \acmConference[SC '21]{SC'21: The International Conference for High Performance Computing, Networking, Storage, and Analysis}{Nov 14--19, 2021}{St.~Louis, MO}
 \acmBooktitle{SC'21: The International Conference for High Performance Computing, Networking, Storage, and Analysis, Nov 14--19, 2021, St.~Louis, MO}

\title{High Performance Uncertainty Quantification with Parallelized Multilevel Markov Chain Monte Carlo}

\author{Linus Seelinger}
\email{mail@linusseelinger.de}
\affiliation{%
  \department{Institute for Scientific Computing}
  \institution{Heidelberg University}
  \city{Heidelberg}
  \country{Germany}
}

\author{Anne Reinarz}
\email{anne.k.reinarz@durham.ac.uk}
\affiliation{%
  \department{Department of Computer Science}
  \institution{Durham University}
  \city{Durham}
  \country{United Kingdom}
}

\author{Leonhard Rannabauer}
\email{rannabau@in.tum.de}
\affiliation{%
  \department{Department of Informatics}
  \institution{Technical University of Munich}
  \city{Garching}
  \country{Germany}
  }

\author{Michael Bader}
\email{bader@in.tum.de}
\affiliation{%
  \department{Department of Informatics}
  \institution{Technical University of Munich}
  \city{Garching}
  \country{Germany}
  }

\author{Peter Bastian}
\email{peter.bastian@iwr.uni-heidelberg.de}
\affiliation{%
  \department{Institute for Scientific Computing}
  \institution{Heidelberg University}
  \city{Heidelberg}
  \country{Germany}
}
  
\author{Robert Scheichl}
\email{r.scheichl@uni-heidelberg.de}
\orcid{0000-0001-8493-4393}
\affiliation{%
  \department{Institute for Applied Mathematics}
  \institution{Heidelberg University}
  \city{Heidelberg}
  \country{Germany}
}
  
%


\begin{teaserfigure}
  \includegraphics[width=\textwidth]{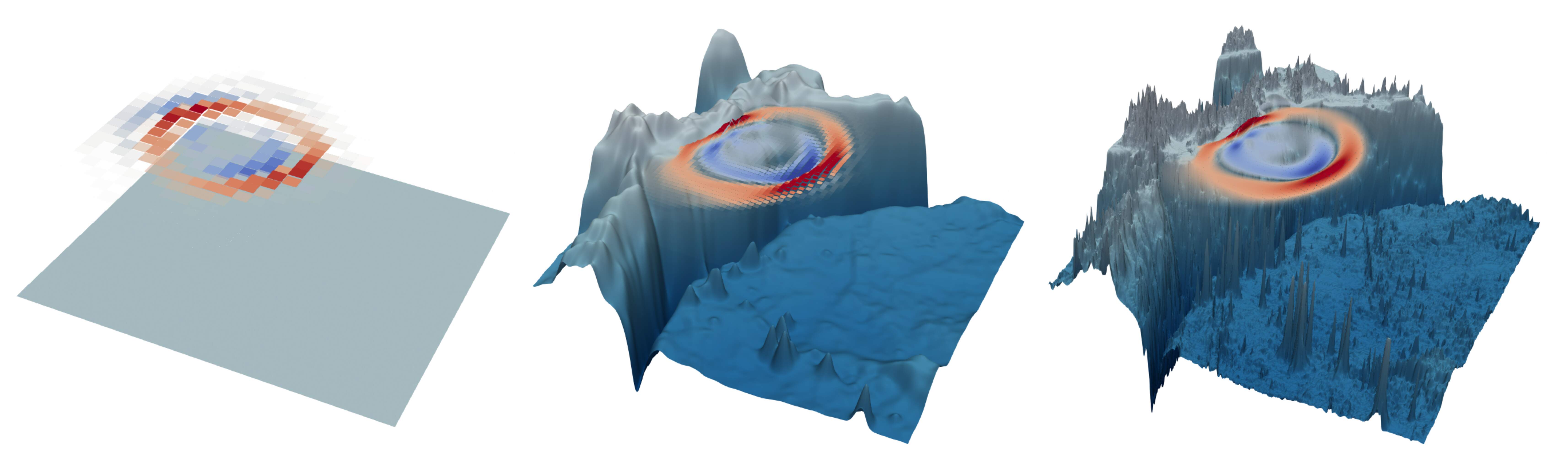}
    \caption{Three levels of discretisation of a tsunami model used in a multilevel Markov chain Monte Carlo method.}
  \label{fig:teaser}
\end{teaserfigure}

\begin{abstract}

Numerical models of complex real-world phenomena often necessitate \gls{HPC}.
Uncertainties increase problem dimensionality further and pose even greater challenges.

We present a parallelization strategy for multilevel Markov chain Monte Carlo,
a state-of-the-art, algorithmically scalable \gls{UQ}
algorithm for Bayesian inverse problems, and a new software framework allowing for
large-scale parallelism across forward model evaluations and the UQ algorithms themselves.
The main scalability challenge presents itself in the form of strong data dependencies introduced
by the MLMCMC method, prohibiting trivial parallelization.

Our software is released as part of the modular and open-source \gls{MUQ},
and can easily be coupled with arbitrary user codes.
We demonstrate it using the \gls{DUNE} and the ExaHyPE Engine.
The latter provides a realistic, large-scale tsunami model in which we identify the source of a tsunami from
buoy-elevation data.

\end{abstract}

\maketitle
\renewcommand{\shortauthors}{L. Seelinger, A. Reinarz et al.}

\newcommand{\revision}[1]{\textcolor{black}{#1}}

\section{Introduction}

Numerical simulation is an established tool driving innovation in many fields of science and engineering.
Numerical solutions of mathematical models, for example in the form of \gls{PDE}, provide a prediction about a real-world process.
More generally, deterministic models map a given set of parameters to a specific set of predicted values.
There are, however, many applications where this approach is insufficient and  a stochastic model is needed:
The exact values of model parameters might be known only up to a certain accuracy,
leading to a corresponding uncertainty in model predictions. The importance of capturing these uncertainties cannot be overstated, since overlooking an unlikely yet dangerous scenario could easily have fatal consequences in real-world applications.

The aforementioned uncertainty in parameters is typically expressed in terms of probability distributions. Finding the distribution of model predictions resulting from stochastic parameter distributions is referred to as a forward \gls{UQ} problem. When in turn real-world measurements, also affected by uncertainty, are available and the underlying stochastic model parameters explaining those measurements are to be quantified, we speak of an inverse \gls{UQ} problem. The resulting parameter distribution is called the posterior.

Here, we focus on inverse \gls{UQ} problems. These are notoriously ill-posed, that is, the solution may lack existence, uniqueness, or continuous dependence on the data. As a result, specialized methods are required for inverse \gls{UQ} problems. There is a wide range of methods available, and they vary signifcantly in how they balance expressivity (in the sense of how much information can be gained) against computational efficiency and against required knowledge about the model. For example, there are highly efficient optimization-based methods such as \cite{KleinMAP} for determining the \gls{MAP} point when derivatives of the data misfit functional are available. On the other hand,  stochastic collocation type methods can work with simple model evaluations
and recover the posterior \cite{Marzouk_StochasticCollocation},
but become inefficient for higher dimensional parameter spaces \cite{SC_Survey}.
\gls{MCMC} type methods in turn also recover the full posterior distribution and in many cases only need simple forward model evaluations, but come at a very high computational cost in terms of numerous model evaluations.

\revision{In this paper, we choose \gls{MCMC} as the ``gold standard'' type of method \cite{LawDataAssimilation} in the sense of
recovering the exact posterior with minimal assumptions on the model. In order to overcome the
enormous computational burden incurred by \gls{MCMC} on costly models,
we employ a multilevel \gls{MCMC} method and present a new parallelization strategy to employ
multilevel \gls{MCMC} on modern \gls{HPC} systems. We then couple our \gls{UQ} method to sophisticated
\gls{PDE} solvers in order to obtain efficient forward models. Since our multilevel \gls{MCMC} method only assumes simple forward model evaluations, no modifications to the solver frameworks are required.  By building efficient model hierarchies, we exploit properties of the forward solvers and show synergies between the \gls{PDE} models and the multilevel \gls{UQ} approach.}
Efficient parallelization is a particular focus of this work since, on the one hand, it is inevitable when solving large-scale \gls{PDE} models, and on the other hand, multilevel \gls{MCMC} introduces data dependencies that make parallelization non-trivial compared to plain-vanilla \gls{MC} methods.

To demonstrate the efficiency of our method \revision{in practical applications}
we use buoy data from coastal Japan to infer the location of initial displacements that led to the Tohoku tsunami in 2011. This method of using DART buoy data to predict tsunamis is commonly used in many early warning systems, such as the one at the Pacific Tsunami Warning Center operated by NOAA in the United States and at the National Tsunami Warning Center. These methods are known to work well for tsunamis which are initiated more than 2-3 wavelengths away from the coast.

We model the propagation of the tsunami by solving the shallow water equations. Our model is capable of modeling wetting and drying so that coastal regions can be included in the simulation \cite{ADER_DG}.
 For the numerical solution of the \gls{PDE}, we apply an \gls{ADER-DG}   implemented in the ExaHyPE framework \cite{ExaHyPE}. The parallelized \gls{MLMCMC} method was implemented in the \gls{MUQ} library \cite{MUQ}\revision{, is publicly available and fully model agnostic. To our knowledge, this is the
 first parallel \gls{MLMCMC} implementation available; a sequential one \cite{lykkegaard2020multilevel}
 is available as part of PyMC3 \cite{pyMC3}.}

We further show parallel scalability of our \gls{UQ} algorithm and implementation by coupling to a model
implemented in the \gls{DUNE}
framework \cite{PDELAB}, at the same time demonstrating how our \gls{UQ} software framework can
be coupled with arbitrary model codes.

\section{Efficient Solution of Inverse Problems via Multilevel MCMC}
\label{sect:MLMCMC}

For a general model $F$ mapping parameters $\theta$ to model predictions $F(\theta)$,
we formulate a Bayesian inverse problem. Its goal is to use the forward model $F$ in
order to infer the distribution of an uncertain model parameter $\theta$ from measurement data $y$:

\begin{equation}
\gls{posteriordens} \coloneqq \gls{posteriordist} = \frac{\gls{likelihood}(\gls{meas}|\gls{param})\gls{priordist}(\gls{param})}{\gls{measdist}}  \propto  \gls{likelihood}(\gls{meas}|\gls{param})\gls{priordist}(\gls{param}).
\label{eq:bayesian_posterior}
\end{equation}

Here, \gls{posteriordist} is the inferred parameter distribution given measurements \gls{meas}.
The likelihood \gls{likelihood} is defined as
the probability of observing the given measurements
if \gls{param} were the true parameter. 
Further, \gls{priordist} is the prior distribution encoding
a priori knowledge about the inferred parameters.
We treat the distribution of measurements \gls{measdist}
as a practically unobtainable scaling factor, as it is irrelevant for \gls{MCMC} methods.
The final target is the mean of a \gls{QOI} $\mathbb{E}_{\gls{posteriordens}}[\gls{qoi}]$ with respect to the
above posterior distribution.

Evaluating the likelihood typically means comparing the model prediction $F(\theta)$
to the measurements \gls{meas}.
For example, assuming Gaussian measurement errors with covariance $\Sigma_f$,
the likelihood will have the distribution $\mathcal{N}(F(\theta), \Sigma_f)$.
In our applications, evaluating $F$ means solving a \gls{PDE},
so this is where the main cost of a Bayesian problem lies.

The main idea of \gls{MCMC} is to generate a Markov chain, carefully designed to have a stationary distribution
matching the posterior distribution we are looking for.
Thus, for a sufficiently large number of steps, the chain will approximately draw samples
from the otherwise inaccessible posterior.
Those samples can in turn be used in a \gls{MC}-like fashion to estimate $\mathbb{E}[Q]$
or other statistics of the posterior.

\Cref{algo:MCMC} shows one of the most common \gls{MCMC} algorithms, namely \gls{MHMCMC} \cite{MHMCMC}.
One of its main advantages is that it only requires a finite number of
direct evaluations of the posterior and thereby of the forward model.
No further information such as model derivatives or adjoints are needed.

\begin{algorithm}
    \SetAlgoLined
    \KwResult{Markov chain $\{\gls{param}^i\}_{i=0}^{\gls{numsamps}}$.}
    Choose starting parameter $\gls{param}^0 \in \mathbb{R}^m$; \\
    \For{$i = 0, ..., \gls{numsamps} - 1$}{
        Draw proposal $\gls{param}'$ from proposal distribution $\gls{propdist}(\gls{param}' | \gls{param}^i)$.\\
        Compute acceptance probability
\[
\alpha(\gls{param}' | \gls{param}^i) = \min{ \left\{ 1, \frac{\gls{posteriordens}(\gls{param}') \gls{propdist}(\gls{param}^i | \gls{param}')}{\gls{posteriordens}(\gls{param}^i) \gls{propdist}(\gls{param}' | \gls{param}_i)} \right\} }.
\]
        Draw a random number $r \in [0,1]$.\\
        \eIf{$r < \alpha(\gls{param}' | \gls{param}^i)$}{
            Accept proposal: $\gls{param}^{i+1} = \gls{param}'$.
        }{
            Reject proposal: $\gls{param}^{i+1} = \gls{param}^i$.
        }
    }
\caption{Metropolis Hastings \gls{MCMC}}
\label{algo:MCMC}
\end{algorithm}

Once we have these samples approximating the posterior, we can apply some post processing to compute
$\mathbb{E}_{\gls{posteriordens}}[Q]$. In practice, since in the \gls{MCMC} algorithm we already compute the
forward model for each sample, we can immediately compute \gls{QOI} samples derived from the model evaluations.

In order to accurately represent a multi-dimensional probability distribution,
a significant number of samples may be required. Just like basic \gls{MC} methods, \cref{algo:MCMC}
requires a full model evaluation per sample. If the model itself is costly, the overall computational effort
can easily become intractable.

Most effort to improve the efficiency of \gls{MHMCMC} goes into finding good proposal
distributions $\gls{propdist}(\gls{param}' | \gls{param}^i)$: \gls{MHMCMC} produces correlated samples
and strongly correlated samples barely contribute
information to the posterior approximation.
Less correlated proposals and high acceptance rates allow to achieve a good approximation with fewer samples.
A wide variety of improved proposals aims to achieve this. Examples include preconditioned Crank-Nicolson \cite{pCNOrig,pCNRevised,pCNGeneralized}, Adaptive
Metropolis \cite{AMMCMC, AMMCMC2}, Hamiltonian MCMC \cite{HamiltonianMCMC},
Dimension-Independent Likelihood-Informed (DILI) MCMC \cite{DILI,MultilevelDILI}, and many others.

Rather than minimizing the number of model evaluations for a single model, \gls{MLMCMC} \cite{SchwabMLMCMC,MLMCMC,MLMCMCRevised} takes a different route:
An entire hierarchy of models is defined, ranging from cheap-to-compute rough approximations to the most accurate, yet expensive, full model.
\gls{MLMCMC} makes no assumptions on what exactly those coarse models could be. For example, we could use suitable \gls{ODE} as
rough approximations of a more complex \gls{PDE} model, as long as model evaluations are sufficiently close.
A more obvious and theoretically supported choice for a level hierarchy is varying mesh width in numerical \gls{PDE} solvers, where
theory typically guarantees that coarser meshes still deliver reasonable approximations.
Note that the fundamental approach is very closely related to \gls{MLMC} methods \cite{MLMC}. However, as we will see later,
there are additional intricacies and benefits to be gained in the \gls{MLMCMC} setting.

The basic idea of \gls{MLMCMC} is to replace the estimation of the expected value of the \gls{QOI}
by a telescoping sum

\begin{equation}
\mathbb{E}_{\gls{posteriordens}_L}[Q_L] = \underbrace{\mathbb{E}_{\gls{posteriordens}_0}[Q_0]}_{\text{Coarse approx.}} + \sum_{l=1}^{L}\underbrace{( \mathbb{E}_{\gls{posteriordens}_l}[Q_l]-\mathbb{E}_{\gls{posteriordens}_{l-1}}[Q_{l-1}] )}_{\text{Corrections}},
\label{eq:telescoping_sum}
\end{equation}

where $Q_l$ denotes the approximation of the \gls{QOI} $Q$ using model $l$ of the model hierarchy.
Working with a hierarchy of models not only affects the \gls{QOI}: Each forward model $F_l$ now
implies its very own likelihood function $\mathcal{L}_l$ and corresponding posterior density
$\gls{posteriordens}_l$ according to the Bayesian inverse problem formulated above.

This reformulation is clearly equivalent, but offers opportunities for significantly reducing
computational cost:
\begin{itemize}
 \item Coarser chains can be used as cheap-to-compute, well informed and nearly uncorrelated proposals for finer chains.
 \item Variance reduction between levels can be exploited:
The coarser corrections in \cref{eq:telescoping_sum} need many samples, but those are cheap to compute;
fine corrections are more expensive per sample, but the variance in those corrections is reduced and
thus only few samples are needed, especially when the models are converging as $l \rightarrow \infty$.
\end{itemize}

An \gls{MLMCMC} method implementing both ideas is shown in \cref{algo:mlmcmc}, as originally introduced
by the authors of \cite{MLMCMC,MLMCMCRevised}.

\begin{algorithm}
\SetAlgoLined
\KwResult{Markov chains $\{\gls{param}^i_l\}_{i=0}^{\gls{numsamps}_l}$ for all levels $l \in \{0, ..., L\}$.}
\caption{Multilevel MCMC}
\label{algo:mlmcmc}

On level 0, run a conventional \gls{MCMC} chain, delivering samples $\gls{param}_0^i$.

\For{level $l = 1, \ldots, L - 1$}{
Choose starting point $\gls{param}_l^0$ with coarse component from
next coarser starting point $\gls{param}_{l-1}^0$.

\For{sample $j = 1, \ldots, N_l$}{

    Given $\theta_l^j$, generate proposal $\theta'_l = \left\{ \begin{array}{c}
    \theta_{l, C}'\\
    \theta_{l, F}
    \end{array} \right\}$ where

    \begin{itemize}
    \item $\theta'_{l, C}$ is drawn from a level $l-1$ chain with \\ subsampling rate $\rho_l$ and

    \item $\theta_{l, F}'$ is drawn from a proposal \\ density $q_l ( \theta_{l, F}' |
    \theta_{l, F}^j)$.
    \end{itemize}

    Compute acceptance probability
    
    \begin{equation*}
    \hspace{-3.5em} \alpha ( \theta'_l, \theta_l^j) = \min \left\{ 1, \frac{\gls{posteriordens}_l ( \theta'_l)
    q_l ( \theta_{l,F}^j | \theta'_{l,F} )}{\gls{posteriordens}_l ( \theta^j_l) q_l (
    \theta'_{l,F} | \theta_{l,F}^j )} \cdot \frac{\gls{posteriordens}_{l - 1} ( \theta^j_{l,
    C}) }{\gls{posteriordens}_{l - 1} (
    \theta'_{l, C}) }
    \right\}. \end{equation*}

    Draw a random number $r \in [0,1]$.\\
    \eIf{$r < \alpha ( \theta'_l, \theta_l^j)$}{
        Accept proposal: $\theta'_l$ as $\theta_l^{j+1}$
    }{
        Reject proposal: $\theta_l^{j+1} = \theta_l^{j}$
    }
}
}
\end{algorithm}

Here we begin with a regular \gls{MCMC} chain for the coarsest level. We then
proceed to generate \gls{MCMC} chains for finer levels, while using samples from coarser chains
as proposals. Since we permit increasing parameter dimensions across levels, the coarse samples $\gls{param}_{l,C}'$ from
level $l-1$ used as proposals for level $l$ may need to be complemented with a fine proposal density $\gls{propdist}_l$
in order to form a proposal $\gls{param}_l'$ of suitable dimension.
Further, compared to \gls{MHMCMC}, the acceptance probability also needs to be adapted to avoid bias
from the coarse proposals.

The efficiency gain of this multilevel algorithm over \gls{MHMCMC} can be analyzed theoretically for
models where \gls{PDE} theory provides approximation error bounds for given cost. In
\cite[Thm. 3.4]{MLMCMC},
the following cost estimate was proven for a Poisson equation model in $d$ dimensions similar to the Poisson model we use later.
It is based on the mean square error of the estimate of $\mathbb{E}_{\gls{posteriordens}}[Q]$,
either directly computed via \gls{MHMCMC} or from the telescoping sum \cref{eq:telescoping_sum} in the \gls{MLMCMC} case.

The computational cost $\gls{cost}_\epsilon$ required to achieve a mean square error below $\epsilon$ is then bounded by
\[ \gls{cost}_\epsilon^{\revision{\text{MCMC}}} \lesssim \epsilon^{-(d+2)-\delta}
\qquad and \qquad
   \gls{cost}_\epsilon^{\revision{\text{MLMCMC}}} \lesssim \epsilon^{-(d+1)-\delta}\]
respectively, where $\delta$ is a model specific constant. In this setting, which should be representative of many \gls{PDE} models with model hierarchies based on mesh width,
the cost for the multilevel method is therefore one order below the single-level one.

The hidden constants in the bounds are proportional to the integrated autocorrelation times of the Markov chains.
On the finest level in the multilevel approach this factor is essentially
reduced to one \cite{MLMCMCRevised}, leading to significant additional computational gains.

It is beyond the scope of this work to also prove the assumptions underlying the theoretical error estimates
for our more complex tsunami model. This would be a particularly challenging task since we do not only rely on mesh
refinement, but change bathymetry data across levels for improved solver performance.
However, since our numerical \gls{PDE} solver should fundamentally exhibit comparable scaling behaviour, we can reasonably expect similar gains
in efficiency by employing the multilevel method in that setting as well.

\section{Models}

We will apply parallelized \gls{MLMCMC} to two different models:
The widely used and cheap to compute Poisson equation modelling single-phase subsurface flow
as well as a more expensive and
realistic Tsunami model. The former serves as a well-known reference, and due to its low cost
is also suitable to gauge the \gls{MLMCMC} parallelization in relatively short scaling experiments.
The latter, due to its more interesting properties, allows us to demonstrate the practical value of
the method, as well as the opportunities in creating model-specific hierarchies.

\subsection{Poisson Model}

In this example, our forward model maps a parameter $\theta$ that models the uncertainty in the diffusion coefficient
to the solution of the Poisson \gls{PDE} evaluated at certain points.
The inverse problem consists in estimating
the underlying parameter $\theta$ from given synthetic data while taking uncertainty into account.

Specifically, we solve the \gls{PDE}

\[\nabla \cdot (\kappa(x,\theta) \nabla u(x,\theta)) = 0 \qquad \mbox{for} \quad x \in \Omega \mbox{ and } \theta \in \Theta, \]

where we choose the domain $\Omega \coloneqq [0,1]^2$ and $\Theta \coloneqq \mathbb{R}^m$. As boundary conditions,
we apply $u(x)=0$ on the left, $u(x)=1$ on the right and natural Neumann boundary conditions at the remainder of
the boundary. We model $log(\kappa)$, the logarithm of the diffusion coefficient,
as a zero-mean Gaussian random field with correlation length $0.15$ and variance 1.
In order to arrive at a finite dimensional
representation of the field, we truncate use a \gls{KL} expansion which we truncate after $m$ terms, i.e.

\[ \log{(\kappa(x,\theta))} \approx \sum_{k=1}^{m} \phi_k(x) \theta_k, \]

where $\phi_1, ..., \phi_m$ are the \gls{KL} modes of largest wave length. \revision{Consequently, $\theta$
is a vector of \gls{KL} coefficients.}
We implement the model in the \gls{DUNE} framework \cite{PDELAB} with a $\mathcal{Q}_1$ Finite Element discretization
on simple structured grids. To form a three-level model hierarchy for our \gls{MLMCMC} method,
we choose mesh widths of $\frac{1}{16}, \frac{1}{64}$ and $\frac{1}{256}$. \revision{Across all three levels,
we choose an identical parameter dimension $m=113$; this specific number carries not much significance
beyond being suitable for implementation
in the circulant embedding \cite{CirculantEmbedding} based random field generator \texttt{dune-randomfield}.}

\begin{figure}
   \includegraphics[width=0.45\columnwidth]{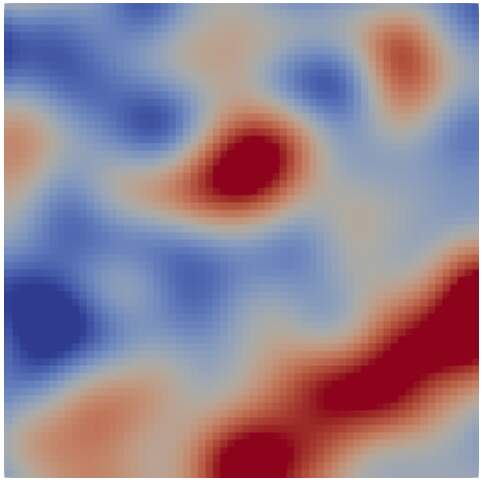}
   \hfill
   \includegraphics[width=0.45\columnwidth]{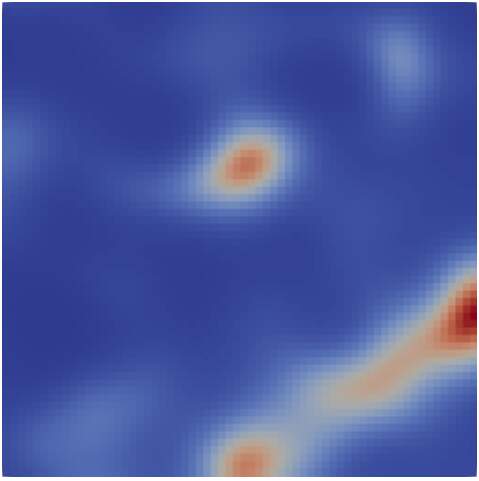}
    \caption{Random field realization $log(\kappa(\cdot, \hat{\theta}))$ and parameter field $\kappa(\cdot, \hat{\theta})$ from random field realization used for synthetic data.}
  \label{fig:field_reference}
\end{figure}

In order to form a Bayesian inverse problem, we generate synthetic data based on
a random field $\kappa(x, \hat{\theta})$, where $\hat{\theta}$ is a fixed sample drawn from $\mathcal{N}(0,I)$
(shown in \cref{fig:field_reference}).

The actual vector of measurements $y$ is then defined by solving the above Poisson problem for $\theta = \hat{\theta}$,
evaluating the solution $u$ at a grid of points:
$\{ \frac{2}{32}, \frac{7}{32}, \frac{13}{32}, \frac{19}{32}, \frac{25}{32}, \frac{3}{32} \}^2$.
Based on that, we define our likelihood $\mathcal{L}(y | \theta)$ to be a Gaussian $\mathcal{N}(F(\theta), \sigma_F^2 I)$ with $\sigma_F = 0.01$.
Complementing it with a Gaussian prior $\pi(\theta) = \mathcal{N}(0,4\revision{I})$, we complete our Bayesian inverse problem.

\revision{Note that, by generating synthetic measurements directly from our forward model, we
commit an 'inverse crime' \cite[p. 179]{ColtonInverseScatteringTheory}.
In realistic applications a model error is inevitable, making it significantly harder to recover
the underlying parameters from data accurately. However, for this problem we intentionally accept this simplification:
Our focus in this case is algorithmic scalability and not a fully realistic setting. Further, verifying the
correctness of the \gls{UQ} method is somewhat easier without a model error.
}

As our \gls{QOI}, we define $Q(\theta)_k = \kappa(x_k, \theta)$ where the $x_k$ form a grid
of width $\frac{1}{32}$. This captures the parameter field we seek and, as necessitated by the telescoping sum in \cref{eq:telescoping_sum},
allows for a consistent dimension in \gls{QOI} even when varying the
parameter dimension across levels.

\subsection{Tsunami Model}

\label{sect:tsunami_model}

Tsunami propagation is typically modelled by some variant of the shallow water equations \cite[e.g.]{Behrens,leveque2011tsunami}, allowing for a simulation in only two dimensions.
The shallow water equations are obtained via depth-averaging of quantities (esp.\ momentum) from the more complicated three-dimensional Navier-Stokes equations, based on the modeling assumption that horizontal length scales are considerably greater than the vertical length scales. 

In this setting, we concentrate on the basic shallow water equations with bathymetry source terms (neglecting friction terms or more advanced models for with non-hydrostatic corrections). 
The resulting equations can be written in first-order hyperbolic form as
\begin{equation}\label{eq-swe}
\frac{\partial}{\partial t}
\begin{pmatrix}
h\\hu\\hv\\ b
\end{pmatrix} + \nabla \cdot
\begin{pmatrix}
hu   &   hv\\
hu^2 & huv\\
huv & hv^2 \\
0 & 0\\
\end{pmatrix}+
\begin{pmatrix}
0\\
hg \, \partial_x (b+h)\\
hg \, \partial_y (b+h)\\
0\\
\end{pmatrix}= 0,
\end{equation}
where $h$ denotes the height of the water column, $(u,v)$ the horizontal flow velocity, $g$  gravity and $b$ denotes the bathymetry. This hyperbolic system of equations is supplemented by a set of suitable initial and boundary values. 

We discretise with an \gls{ADER-DG} method as proposed in \cite{Dumbser}. It is essentially a predictor-corrector scheme. A high-order solution is found element-locally and then corrected to take into account neighbors by solving Riemann problems along element interfaces. To resolve known high-order issues such as the Gibbs phenomenon a corresponding a-posteriori finite volume sub-cell limiter is applied \cite{FV_Limiter}. This limiter detects and revokes problematic \gls{ADER-DG} solution candidates and recomputes them with a robust Finite-Volume scheme (cf.~\cite{rannabauer_enviro,ADER_DG} for details), following the approach by LeVeque et al.~\cite{leveque2011tsunami}. 
At coastlines the schemes relies entirely on the Finite-Volume limiter, to correctly treat inundation. 

As a large-scale example, we invert data from the Tohoku tsunami, which occured subsequent to an earthquake in the Japan trench in 2011.
We assume that the only significant sources of the tsunami are the displacements of the sea floor.
In order to initialise the tsunami, we can impose the displacements as an instantaneous deformation of the bathymetry in the resting-lake case -- 
compare respective modeling approaches by Saito et al.~\cite{saito_deformation} or Madden et al.~\cite{madden_ascete}.
By keeping the water column constant, the change of the bathymetry is directly translated to the sea surface and generates the tsunami.
Gravity, as the main acting force, initiates the propagation of the wave.
The tsunami then evolves as a gravity wave.
\begin{figure}
    \centering
   \includegraphics[width=0.5\columnwidth]{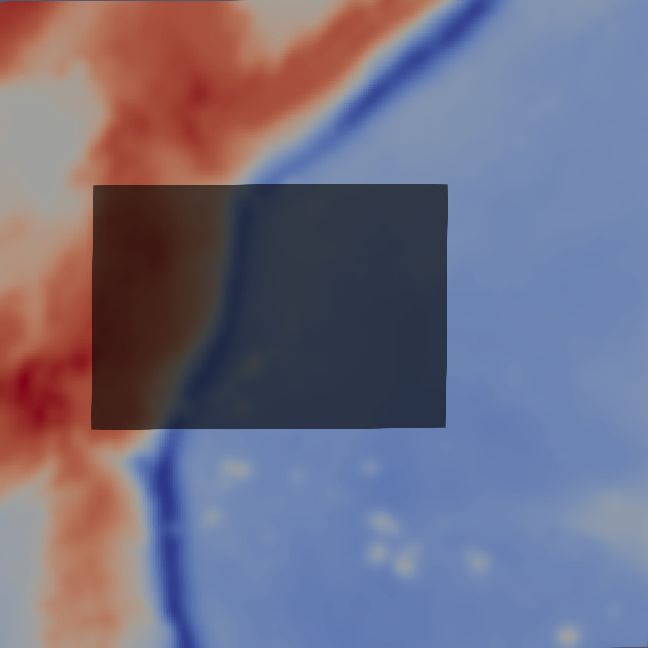}
    \caption{Bathymetry for the full computational domain with darker rectangle denoting parameter values in the prior.}
  \label{fig:prior}
\end{figure}
%
As reference solution for the initial displacements of the ocean floor we use respective simulation results provided by Galvez et al.~\cite{Galvez}.
The bathymetry data has been obtained from GEBCO \footnote{\url{https://www.gebco.net/}}.

Our goal is to obtain the parameters describing the initial displacements from the data of two available buoys located near the Japanese coast. Some of these parameters are reasonably well-known already, these include location of the hypocenter, length, width, and to some extent depth. Other parameters such as uplift are more difficult to estimate. In these tests we estimate the location of the initial displacement. The prior cuts off all parameters which would lead to an initial displacement which is too close to the domain boundary. Figure \ref{fig:prior} shows the cut-off values used. \revision{Some parameters may lead to unstable models, e.g. a parameter which initialises the tsunami on dry land, in this case we have treated the parameter as unphysical and assigned an almost zero likelihood.}

\begin{figure}
   \includegraphics[width=\columnwidth]{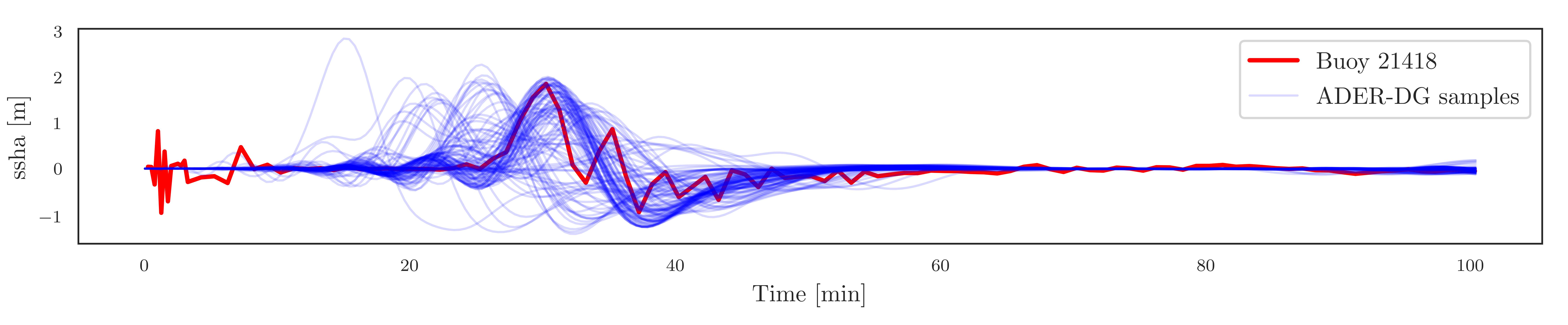}
    \includegraphics[width=\columnwidth]{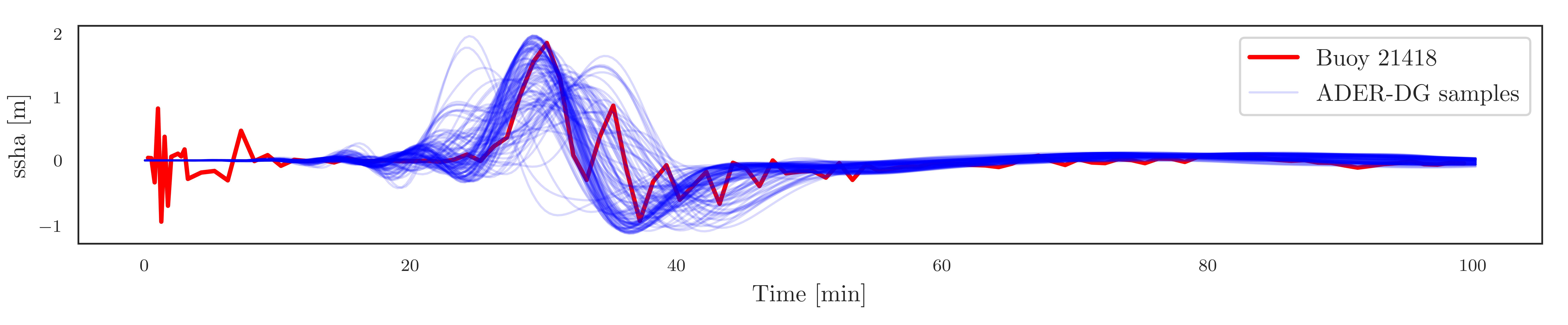}
    \caption{Plot of Sea surface height anomaly [ssha] for samples taken from level 0 (top) and 1 (bottom) compared to NDBC data at buoy 21418.}
  \label{fig:probes_18}
\end{figure}
\begin{figure}
   \includegraphics[width=\columnwidth]{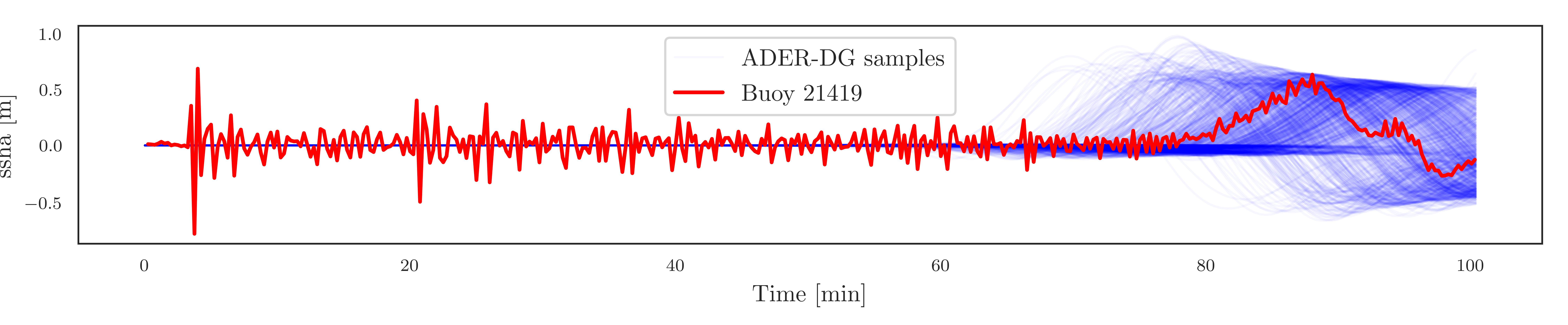}
    \includegraphics[width=\columnwidth]{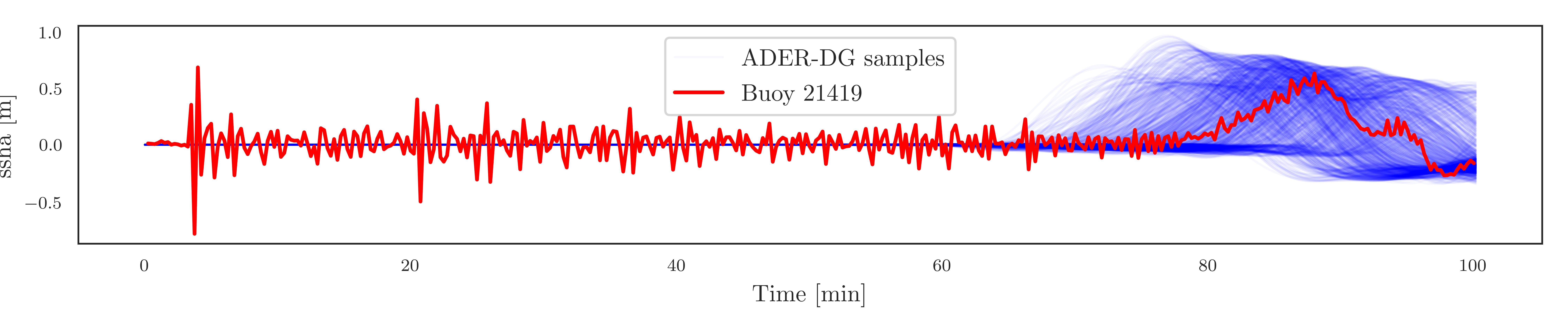}
    \caption{Plot of Sea surface height anomaly [ssha] for samples taken from level 0 (top) and 1 (bottom) compared to NDBC data at buoy 21419.}
  \label{fig:probes_19}
\end{figure}
\begin{table}
\begin{center}
 \begin{tabular}{||c c c c||} 
 \hline
     $\mu$ & \multicolumn{3}{c||}{$\Sigma$} \\ [0.5ex]
           & l=0 & l=1 & l=2\\[0.5ex]
 \hline\hline
 1.85232 & 0.15 & 0.1 & 0.1 \\ 
 \hline
 0.6368 & 0.15 & 0.1 & 0.1 \\
 \hline
 30.23 & 2.5 & 1.5 & 0.75 \\
 \hline
 87.98 & 2.5 & 1.5 & 0.75 \\
 \hline
\end{tabular}
\end{center}
\caption{Mean $\mu$ and covariance $\Sigma$ for all three levels.}
\label{tbl:mu}
\end{table}
In order to compute a likelihood of a given set of parameters given the simulation results we use a weighted average of the maximal wave height and the time at which it is reached. The likelihood is given by a normal distribution
$\mathcal{N}\left(\mu, \Sigma \right)$
with mean $\mu$ given by maximum waveheight $\max\{h\}$ and the time $t$ at which it is reached for the the two DART buoys 21418 and 21419 \footnote{This data can be obtained from NDBC \url{https://www.ndbc.noaa.gov/}}. The covariance matrix $\Sigma$ depends on the level, but not the probe point. \Cref{fig:probes_18,fig:probes_19} shows the data and samples of level $0$ and $1$.
Table \ref{tbl:mu} gives the values of $\mu$ and the diagonal entries of $\Sigma$ for all three levels. Alternative likelihood functions, such as a quadratic average of multiple buoys could also be used, see e.g. \cite{Behrens}.

We set up a sequence of three models with increasing accuracy shown as in Figure \ref{fig:teaser}. In the first model bathymetry is approximated only by a depth average over the entire domain. Since no calculations of wetting and drying are needed this model is computed purely with a DG method of order 2. The second  and third models further include a finite volume subcell limiter allowing for wetting and drying. The second model uses smoothed bathymetry data and the third uses the full bathymetry data. The main advantage of using smoothed data is that the FV subcell limiter is needed in fewer cells. This model hierarchy demonstrates that not only mesh refinement and coarsening, but also model specific optimisations can be used to exploit multilevel \gls{MCMC}.

\revision{It should be noted that the limiter, which works on a locally refined grid, leads to a varying computational load per sample. The number of degrees of freedom for each model are given in Table \ref{tbl:dof} for the sample calculated with the parameters $(0,0)$.}

\begin{table}
\begin{center}
 \begin{tabular}{||c c c c c c||} 
 \hline
 \revision{level} &  \revision{order}        & \revision{limiter} & \revision{h} &\revision{ \# timesteps} & \revision{DOF updates} \\[0.5ex]
 \hline\hline
 \revision{0} & \revision{2} & \revision{no} & \revision{$1/25$} & \revision{$98$} & \revision{$2.4\cdot 10^5$} \\[0.5ex]
\hline
\revision{1} & \revision{2} & \revision{yes} & \revision{$1/79$} & \revision{$306$} & \revision{$9.4\cdot 10^6$} \\[0.5ex]
 \hline
 \revision{2} & \revision{2} & \revision{yes} & \revision{$1/241$} & \revision{$932$} & \revision{$2.7\cdot 10^8$} \\[0.5ex]
 \hline
\end{tabular}
\end{center}
\caption{\revision{Polynomial order, limiter status, mesh width ($h$), number of timesteps and number of degree of freedom (DOF) updates for the three models.}}
\label{tbl:dof}
\end{table}

\section{Parallel MLMCMC Implementation}

In this section we describe the new, highly scalable parallel implementation of \gls{MLMCMC} used for the experiments in this paper.
While parallelization in classical \gls{MC} is trivial due to the independence of the samples, 
\gls{MCMC} introduces data dependency through proposals depending on the previous step.
In the case of \gls{MLMCMC} as in \cite{MLMCMC,MLMCMCRevised}, we use coarser chain samples as proposals, which also
introduces data dependency between levels. There are, however, multiple opportunities for
parallelizing \gls{MLMCMC}:

\begin{itemize}
 \item \textbf{Models}: The forward models themselves may be parallelized. In fact, for large models like the
       tsunami model we introduce in \cref{sect:tsunami_model},
       that is inevitable anyway due to memory constraints.
 \item  \textbf{Chains:} Instead of running a single Markov chain, multiple chains can be run in parallel
       and their samples combined. It is beneficial to not purely rely on this approach though
       since each chain requires a burn-in phase.
 \item  \textbf{Levels:} Contributions to the multilevel telescoping sum in \cref{eq:telescoping_sum} can
       be evaluated in parallel.
\end{itemize}

Exploiting all those clearly introduces significant technical complexity.
Therefore we provide our implementation as part of the \gls{MUQ} C++ library \cite{MUQ}.
The main goals of this implementation are:

\begin{itemize}
 \item \textbf{Parallelism:} All of the above levels of parallelism are supported.
 \item \textbf{Simple user interface:} We hide the intricate details of communication 
       from the user. The algorithm can be tweaked, but to get started the defaults suffice.
 \item \textbf{Model-agnosticity:} The \gls{MLMCMC} algorithm as detailed in \cref{sect:MLMCMC}
       only requires simple forward evaluations of the model. This theoretically allows
       coupling to arbitrary forward models without any need for, e.g., derivatives of the model map.
       We retain this model agnosticity in the sense that any forward model that can be called (possibly through wrappers)
       from C++ can be used.
 \item  \textbf{Modularity:} \gls{MUQ} is, from the ground up, designed as a modular framework.
       Its modularity is closely modelled after the respective mathematical objects.
       We extend this concept to our parallel \gls{MLMCMC} implementation by
       building on top of \gls{MUQ}'s existing \gls{MCMC} stack and,
       as detailed in the following, constructing modular parallel units.
\end{itemize}

\subsection{Model interface}

\gls{MUQ} provides an abstract interface for sampling algorithms, including \gls{MCMC} type methods.
In its most basic form, a model can be provided by implementing an \texttt{AbstractSamplingProblem}
(see \cref{fig:UML_SamplingProblem}).
Our parallel \gls{MLMCMC} implementation supports the same model interface,
allowing to quickly use various \gls{UQ} methods on a single model implementation.
The interface requires the implementation of the density to sample from; in case of our Bayesian inverse problem,
that is the posterior density in \cref{eq:bayesian_posterior}.
Further, a \gls{QOI} may be provided. This is implemented as a separate function call, since
discarded samples in \gls{MCMC} do not contribute to the \gls{QOI} computation, and potentially
expensive evaluations can thus be skipped.

\begin{figure}[h]
\centering
\begin{tikzpicture}[scale=.8, transform shape]
\umlclass{AbstractSamplingProblem}{
}{
LogDensity(SamplingState state) : double\\
QOI() : Eigen::VectorXd
}
\end{tikzpicture}
\caption{\texttt{AbstractSamplingProblem} interface.}
\label{fig:UML_SamplingProblem}
\end{figure}

A single implementation of this interface is already sufficient to apply various \gls{MCMC} type methods provided by \gls{MUQ}.
In the multilevel case, however, we clearly need to provide a hierarchy of posterior densities
implying a hierarchy of models. Further, proposal densities and subsampling strategies for
drawing from coarser chains can be chosen.
Simply passing a fixed list of models may lead to various complications and performance issues
in a parallel settings. Therefore, we provide a factory type interface as in \cref{fig:UML_MIComponentFactory}.

\begin{figure}[h]
\centering
\begin{tikzpicture}[scale=.8, transform shape]
\umlclass[]{MIComponentFactory}{
}{
SamplingProblem(index : MultiIndex) : AbstractSamplingProblem\\
FinestIndex() : MultiIndex\\
CoarseProposal(index : MultiIndex, coarseProblem : AbstractSamplingProblem,\\
\hskip 2em coarseChain : SingleChainMCMC) : MCMCProposal\\
Proposal(index : MultiIndex, samplingProblem: AbstractSamplingProblem)\\
\hskip 2em : MCMCProposal\\
Interpolation(index : MultiIndex) : MIInterpolation\\
StartingPoint(index : MultiIndex) : Eigen::VectorXd
}
\umlclass[above=1cm of MIComponentFactory]{ParallelizableMIComponentFactory}{
}{
SetComm(comm : parcer::Communicator) : void
}
\umlinherit[]{ParallelizableMIComponentFactory}{MIComponentFactory}
\end{tikzpicture}
\caption{\texttt{ParallelizableMIComponentFactory} interface.}
\label{fig:UML_MIComponentFactory}
\end{figure}

When implementing the \texttt{MIComponentFactory} interface, the \texttt{SamplingProblem} method is to return
a model for a given model index, while \texttt{FinestIndex} specifies the index of the finest model
the user provides, corresponding to $L$ in \cref{algo:mlmcmc}.
\texttt{CoarseProposal} specifies how proposals are being drawn from coarser chains,
and \texttt{Interpolation} determines how these are combined with a finer chain proposal
as defined in \texttt{Proposal}.

In order to support parallel models,
the user may modify \texttt{ParallelizableMIComponentFactory}, which receives an
MPI communicator. The user can pass a \texttt{parcer} communicator to the model, this is essentially just an MPI C++ wrapper used in \gls{MUQ}.

Note that, while in this paper we entirely focus on the multilevel case, the implementation
actually supports a generalization to a multiindex method. Therefore, components are
named \texttt{MI} for multiindex rather than \texttt{ML}, and the model hierarchy
is indexed by a more general multiindex structure rather than an integer index.

\subsection{Internal architecture}
Our parallel process layout is shown in \cref{fig:ProcessLayout}. We define the following roles:

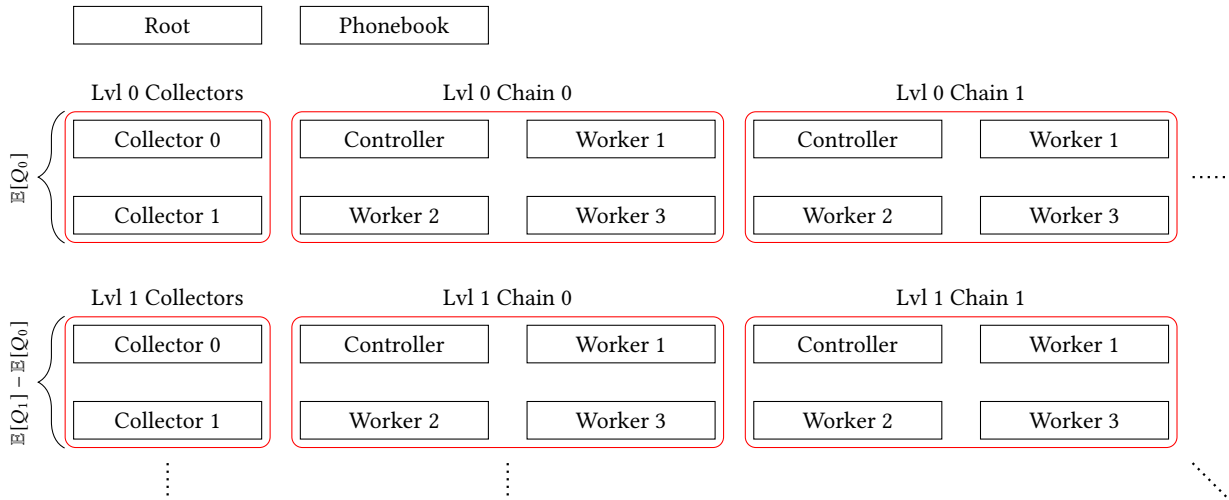
\begin{figure*}[ht]
\centering
\scalebox{1} {
\begin{tikzpicture}[scale=1, minimum width=2.5cm, minimum height=.5cm, transform shape, node distance=.5cm]

\node[draw,label=center:Root] (Root) {};
\node[draw,right=of Root,label=center:Phonebook] (Phonebook) {};

\node[draw,below=1cm of Root,label=center:Collector 0] (Collector_lvl0) {};
\node[draw,below=of Collector_lvl0,label=center:Collector 1] (Collector_lvl0_1) {};
\node (box) [draw=red,rounded corners,fit = (Collector_lvl0) (Collector_lvl0_1), label=Lvl 0 Collectors] (Collector_lvl0_box) {};

\draw [decorate,decoration={brace,amplitude=10pt}]
(Collector_lvl0_box.south west) -- (Collector_lvl0_box.north west)  node [black,midway,xshift=-0.6cm,rotate=90]
{\footnotesize $\mathbb{E}[\gls{qoi}_0]$};

\node[draw,below=1.2cm of Collector_lvl0_1,label=center:Collector 0] (Collector_lvl1) {};
\node[draw,below=of Collector_lvl1,label=center:Collector 1] (Collector_lvl1_1) {};
\node (box) [draw=red,rounded corners,fit = (Collector_lvl1) (Collector_lvl1_1), label=Lvl 1 Collectors] (Collector_lvl1_box) {};
\draw [decorate,decoration={brace,amplitude=10pt}]
(Collector_lvl1_box.south west) -- (Collector_lvl1_box.north west)  node [black,midway,xshift=-0.6cm,rotate=90]
{\footnotesize $\mathbb{E}[\gls{qoi}_1] - \mathbb{E}[\gls{qoi}_0]$};

\draw[-,dotted,line width=0.3mm] (Collector_lvl1_box.south)+(0,-.2) -- +(0,-0.7);


\node[draw,right=of Collector_lvl0,label=center:Controller] (Controller_lvl0_n0) {};
\node[draw,right=of Controller_lvl0_n0,label=center:Worker 1] (Worker_lvl0_n0_w1) {};
\node[draw,below=of Controller_lvl0_n0,label=center:Worker 2] (Worker_lvl0_n0_w2) {};
\node[draw,right=of Worker_lvl0_n0_w2,label=center:Worker 3] (Worker_lvl0_n0_w3) {};
\node (box) [draw=red,rounded corners,fit = (Controller_lvl0_n0) (Worker_lvl0_n0_w3), label=Lvl 0 Chain 0] {};

\node[draw,right=of Worker_lvl0_n0_w1,label=center:Controller] (Controller_lvl0_n1) {};
\node[draw,right=of Controller_lvl0_n1,label=center:Worker 1] (Worker_lvl0_n1_w1) {};
\node[draw,below=of Controller_lvl0_n1,label=center:Worker 2] (Worker_lvl0_n1_w2) {};
\node[draw,right=of Worker_lvl0_n1_w2,label=center:Worker 3] (Worker_lvl0_n1_w3) {};
\node (box) [draw=red,rounded corners,fit = (Controller_lvl0_n1) (Worker_lvl0_n1_w3), label=Lvl 0 Chain 1] (Workgroup_lvl0_n1_box) {};

\draw[-,dotted,line width=0.3mm] (Workgroup_lvl0_n1_box.east)+(.2,0) -- +(.7,0);

\node[draw,right=of Collector_lvl1,label=center:Controller] (Controller_lvl1_n0) {};
\node[draw,right=of Controller_lvl1_n0,label=center:Worker 1] (Worker_lvl1_n0_w1) {};
\node[draw,below=of Controller_lvl1_n0,label=center:Worker 2] (Worker_lvl1_n0_w2) {};
\node[draw,right=of Worker_lvl1_n0_w2,label=center:Worker 3] (Worker_lvl1_n0_w3) {};
\node (box) [draw=red,rounded corners,fit = (Controller_lvl1_n0) (Worker_lvl1_n0_w3), label=Lvl 1 Chain 0] (Workgroup_lvl1_n0_box) {};

\draw[-,dotted,line width=0.3mm] (Workgroup_lvl1_n0_box.south)+(0,-.2) -- +(0,-0.7);

\node[draw,right=of Worker_lvl1_n0_w1,label=center:Controller] (Controller_lvl1_n1) {};
\node[draw,right=of Controller_lvl1_n1,label=center:Worker 1] (Worker_lvl1_n1_w1) {};
\node[draw,below=of Controller_lvl1_n1,label=center:Worker 2] (Worker_lvl1_n1_w2) {};
\node[draw,right=of Worker_lvl1_n1_w2,label=center:Worker 3] (Worker_lvl1_n1_w3) {};
\node (box) [draw=red,rounded corners,fit = (Controller_lvl1_n1) (Worker_lvl1_n1_w3), label=Lvl 1 Chain 1] (Workgroup_lvl1_n1_box) {};

\draw[-,dotted,line width=0.3mm] (Workgroup_lvl1_n1_box.south east)+(.2,-.2) -- +(.7,-.7);


\end{tikzpicture}
}
\caption{Parallel process layout.}
\label{fig:ProcessLayout}
\end{figure*}
\begin{itemize}
 \item \textbf{Fixed roles:} These are assigned to specific processes at the start of the parallel method. All other processes wait to be assigned a dynamic role.

\begin{itemize}
\item \textbf{Root:} This process is responsible for launching the parallel method, assigning tasks to other processes and
  requesting collectors to begin collecting a certain number of \gls{MCMC} samples. It is also the best place for users
  to implement custom (possibly adaptive) sampling strategies.
\item \textbf{Phonebook:} The phonebook tracks what dynamic roles processes are currently assigned to.
Most importantly, it tracks which chains are currently sampling and which ones hold new samples ready to be picked up by other processes. Further, the phonebook can infer from that the computational load on a given level, since
the relation between requested samples and assigned resources is available here. Therefore,
it is also the key component in dynamic load balancing across levels and chains.
\end{itemize}
\item \textbf{Dynamic roles:} These roles may be assigned and reassigned at any time. In particular, this permits dynamic load balancing and possibly more advanced sampling strategies.

Workers and controllers solve forward models, where workers share the load of running a single model evaluation and controllers additionally run the inherently sequential MCMC chains; consequently, they are assigned (and, in case of dynamic load balancing, reassigned) synchronously.
This is faciliated by \gls{MPI} subcommunicators, which are passed through to and should be used by the user's model.
\begin{itemize}
\item \textbf{Worker:} Workers are responsible for running the user's forward model, specifically the user's implementations of \texttt{AbstractSamplingProblem}.
Mathematically, they provide parallelized evaluations of the posterior and quantitiy of interest on a given level for a given parameter.
They listen to their respective controller's
\texttt{ParallelAbstractSamplingProblem} to signal the beginning of a evaluation for a specific parameter \gls{param}. Once the signal arrives, they execute the user-implemented \texttt{LogDensity} method.
Since all workers of a work group are called synchronously, the user can easily supply models assuming lock step parallelism.
\item \textbf{Controller:} Each controller is responsible for running a multilevel \gls{MCMC} chain according to \cref{algo:mlmcmc}. Specifically, a controller on level $l$ contains a chain on level $l$ and one on $l-1$,
to compute the level $l$ correction of the telescoping sum \cref{eq:telescoping_sum} (with the obvious exception of the coarsest level 0). An instance of \texttt{ParallelAbstractSamplingProblem} is set up for each model needed.
It provides an intermediate layer between instances of the user-implemented \texttt{AbstractSamplingProblem} instances running on multiple workers and the controller, allowing to transparently distribute model executions.
As a result, neither the user nor the inherently sequential \gls{MLMCMC} chains
need to concern themselves with synchronizing worker processes to run forward models in lock step.

The chains themselves are implemented using existing \gls{MUQ} components: They are \texttt{SingleChainMCMC} instances
with \texttt{MCMCKernel} implementations matching the acceptance probability of \cref{algo:mlmcmc}. Drawing samples from coarser chains as proposals is implemented with an \texttt{MCMCProposal} requesting coarser samples from other controllers via the phonebook process.
\item \textbf{Collector:} Collectors request samples from controllers via the phonebook in order to compute terms of the telescoping sum \cref{eq:telescoping_sum}. Multiple collectors may be responsible for a single level. Together, they hold a \texttt{DistributedCollection}, an
existing class in \gls{MUQ} for storing and computing statistics on samples in a parallel system.
\end{itemize}

\end{itemize}

Note that this architecture is defined on process level (more specifically, in terms of \gls{MPI} ranks).
Thread-level parallelism can easily be exploited by worker processes. In fact we make use of this through \gls{TBB} in our \gls{ExaHyPE} Tsunami model,
since \gls{ExaHyPE} exhibits better performance characteristics with few \gls{MPI} ranks per node each making use of several threads.

In order to make the parallel architecture as modular as sequential \gls{MUQ} code, each of the above roles provide an \gls{MPI} interface based on requests mimicking function calls.
This allows recombining the parallel components in analogy to object orientation in order to implement other algorithms as well.
For example, work groups as introduced above are based on a \texttt{ParallelAbstractSamplingProblem}, which in turn can be used
to easily employ any sequential sampling algorithm on a parallelized model. Likewise, the phonebook could be swapped out for an alternative implementation with identical \gls{MPI} interface, allowing for alternative load balancing strategies.

\subsection{Load balancing}

Data dependencies in \gls{MLMCMC} (see \cref{algo:mlmcmc}) introduce a load balancing problem, since coarser chains need
to provide proposals to finer ones only until the desired number of fine samples is computed. Estimating
the ideal distribution of computational resources across levels is far from trivial or outright impossible in realistic applications, especially when adaptively determining the number of samples per level.

\begin{figure}
  \includegraphics[width=\columnwidth]{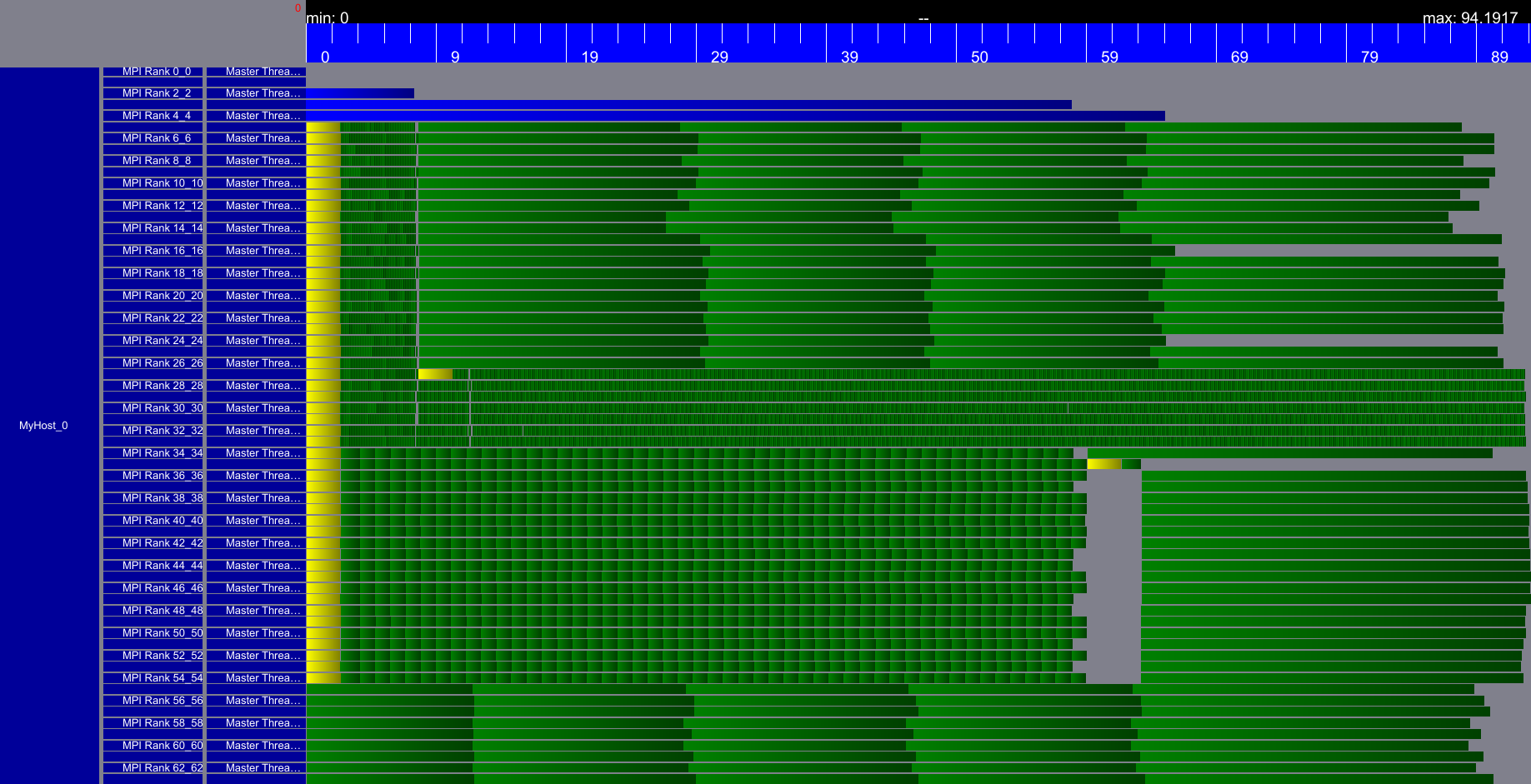}
  \caption{Dynamic load balancing in parallel \gls{MLMCMC}. Run time is on the horizontal axis,
           while process indices are on the vertical. Green boxes indicate model evaluations, while
           yellow boxes indicate chains' burnin phases. In this example, model evaluations
           clearly differ strongly in run time.}
  \label{fig:load_balancing}
\end{figure}

Therefore, our parallel \gls{MLMCMC} implementation provides a load balancing mechanism to reassign
worker processes to different tasks once samples on another level are more critical to runtime.
\Cref{fig:load_balancing} illustrates this load balancing mechanism for a small test run.

Load balancing is implemented as part of the phonebook rank, since it keeps track of how samples are passed around.
Levels with low load are detected when samples on that level are provided but not quickly picked up,
while a high load is in turn detected when sample requests remain queued. Unanswered sample requests originating
from other chains are given a higher impact than requests originating from collector processes, since
the first case implies chains waiting and therefore bad machine utilization.

Models may have strongly varying run times. A new group of processes assigned to a certain level
only reduces that level's load once it actually provides its first sample.
This implies the danger of reassigning tasks too frequently or too infrequently. In order to stabilize the
load balancer, the respective model run times are inferred by the phonebook process by the frequency of samples provided.
Based on that, scheduling will only take place at the time scale of the respective model evaluations.

Note that this load balancer is unaware of the specific types of proposals or \gls{MCMC} kernels
being executed. As a result, it can, for example, also be applied in the \gls{MLMC} setting.
Also, all model specific tuning parameters are determined dynamically, so no user intervention is required.

\section{Numerical Results}

\subsection{Poisson application and scalability}

\begin{figure}
   \includegraphics[width=0.45\columnwidth]{images/perm_field_reference.png.conv.jpg}
   \hfill
   \includegraphics[width=0.45\columnwidth]{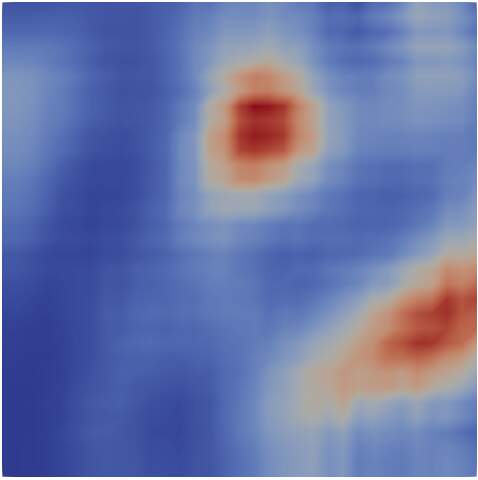}
    \caption{Synthetic ``true'' field (left) and expected value of multilevel estimator (right).}
  \label{fig:ml_mean}
\end{figure}
\begin{table}
\begin{center}
 \begin{tabular}{||c c S[table-format=5] S[table-format=3.2] c S[table-format=3.2] c||}
 \hline
               &       &      &       &          &          & $\mathbb{V}[Q_0]$ or\\
     level $l$ & $h_l$ & $\revision{\text{DOFs}}$ & $t_l$ \revision{[ms]} & $\rho_l$ & $\tau_l$ & $\mathbb{V}[Q_l - Q_{l-1}]$\\
 \hline
 \hline
 0 & $\frac{1}{16}$  & 289   & 3.35   & 206 & 137.3 & $1.501 \times 10^{-1}$\\
 1 & $\frac{1}{64}$  & 4225  & 45.64  & 17 & 11.2 &    $1.121 \times 10^{-3}$\\
 2 & $\frac{1}{256}$ & 66049 & 931.81 & 0 & 1.05 &     $4.165 \times 10^{-5}$\\
 \hline
\end{tabular}
\end{center}
\caption{Multilevel properties of Poisson application.
For each level $l$ of mesh width $h_l$ \revision{with associated degrees of freedom (DOFs)}, we show the computational cost $t_l$ and chosen subsampling rate $\rho_l$.
Due to high dimension of \gls{QOI} in this setting, we only show integrated autocorrelation time \revision{$\tau_l$} and variance for an
single representative component of $Q$.
}
\label{tbl:poisson_ml_setup}
\end{table}
\revision{In order to fully specify the \gls{MLMCMC} algorithm for the given problem,
it is enough to set a Gaussian proposal on the coarsest level.
We choose $\mathcal{N}(0, 3 I)$ in order to roughly match the prior.
Since we have identical parameter dimensions across levels, no fine level proposals are needed.}

The \gls{MLMCMC} method run with $10^4$, $10^3$ and $10^2$ samples on levels 0, 1 and 2 exhibits properties detailed in
in \cref{tbl:poisson_ml_setup}
and captures the main features of the parameter field underlying our synthetic data (see \cref{fig:ml_mean}).
Clearly some higher frequency detail is not recovered. This, however, is expected due to the
limited number of \gls{KL} modes we include in our parameter space. Note that we can only recover this
up to a scaling factor, since the solution is only determined by the parameter field up to a factor.
Here, the choice of prior essentially determines the scaling of the solution we observe.

For a more detailed analysis of Bayesian inverse problems based on Poisson equation in \gls{MLMCMC},
we refer to the original \gls{MLMCMC} publication \cite{MLMCMC,MLMCMCRevised}.

In order to investigate parallel scalability of our \gls{MLMCMC} implementation,
we conduct weak and strong scaling experiments on the BwForCluster MLS\&WISO Production \gls{HPC} system.
The partition we used consists of nodes with two 16-core Intel Xeon E5-2630v3 CPUs and 64 gigabytes of memory.

As forward model, we use the Poisson model, since its low computational demand allows us to stress
the parallelized \gls{MLMCMC} algorithm itself by running a large number of chains and samples.
We use the same inverse problem detailed above, even though the particular inverse problem does not
affect the algorithm's communication patterns and therefore parallel scalability.

\begin{figure}
  \includegraphics[width=0.9\columnwidth]{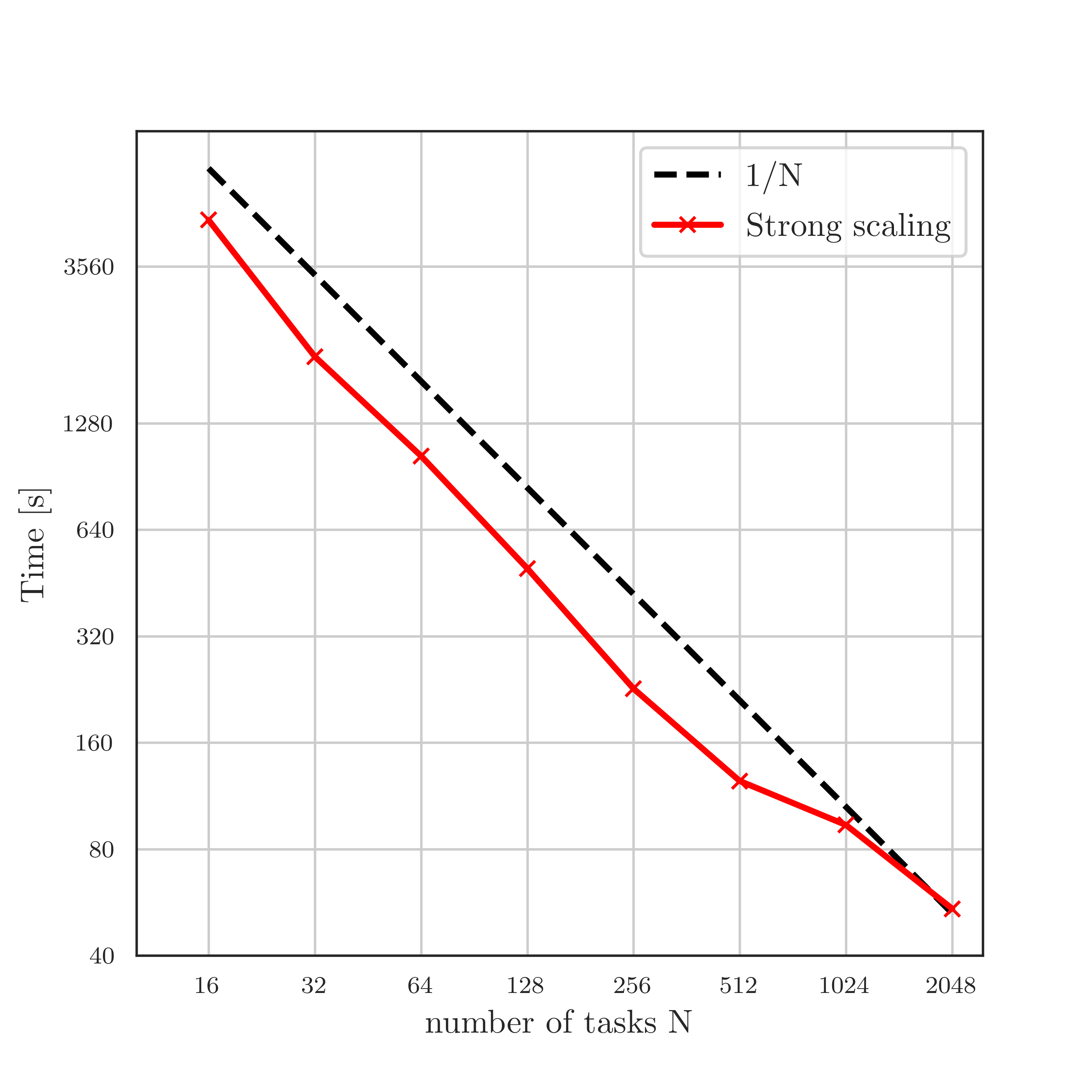}
  \caption{Scalability of the Poisson model problem for $10^4$, $10^3$ and $10^2$ samples on levels 0, 1 and 2 respectively. \revision{Subsampling rates etc. are chosen according to \cref{tbl:poisson_ml_setup}}. The problem setup remains constant as the number of processors is increased.}
  \label{fig:ss}
\end{figure}
For the strong scaling setup, we draw $10^4$, $10^3$ and $10^2$ samples on levels 0, 1 and 2 respectively.
We further set subsampling rates according to \cref{tbl:poisson_ml_setup}, and enable dynamic load balancing.
As the timing results in \cref{fig:ss} show, we achieve linear speedup until
relatively large burnin phases and suboptimal load balancing due to few samples per chain occur.

Technically the observed speedup even slightly exceeds linear. That is simply due to the fact that a fixed number
of the processes is reserved for book keeping tasks (i.e. the root, phonebook and collector processes).
As a result, for increased number of processes, a larger fraction contributes to parallel speedup
by generating samples.

\begin{figure}
  \includegraphics[width=0.9\columnwidth]{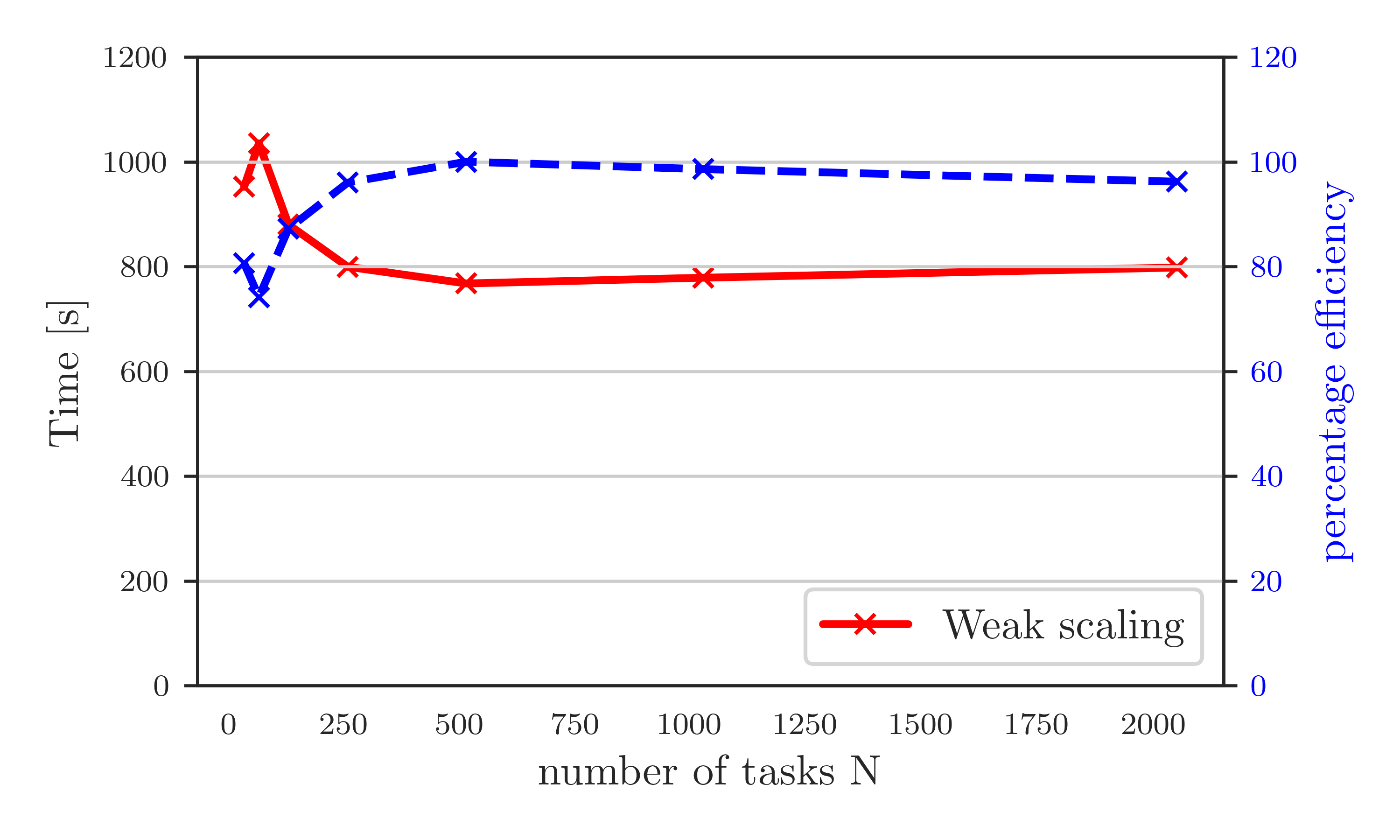}
  \caption{Weak scalability \revision{and parallel efficiency} of the Poisson model problem. At $64$ cores $10^4$, $10^3$ and $10^2$ samples are computed on levels 0, 1 and 2 respectively. The number of samples is modifed linearly with the number of processors.}
  \label{fig:ws}
\end{figure}
In our weak scaling test, we begin with the same setup as in the strong scaling setting.
In particular, we again choose to compute $10^4$, $10^3$ and $10^2$ samples on levels 0, 1 and 2,
and solve this problem using 64 processes. We then expand to a range from $32$ to $1024$ processes
while scaling the number of samples on each level linearly in accordance with the number of processes.

\revision{
The parallel efficiency (given in blue in Figure \ref{fig:ws}) is measured here as $\frac{t_\text{ref}}{ t_N } \cdot 100\% $, where $t_\text{ref}$ is the quickest time taken over all runs and $t_N$ is the time taken on $N$ ranks. The initial increase to over $100\%$ efficiency is due to the overhead of phonebook and collector ranks.}
We achieve fairly consistent results of up to 80 seconds of total run time except for the largest run.
The latter is a very extreme scenario though, since the extremely short run time of the coarsest model
leads to a significant load on the communication infrastructure. We therefore consider it reasonable
to assume that exceeding the ideal range should only occur for significantly larger numbers of processes
in more realistic applications.

\subsection{Tsunami application}

\begin{figure}
   \includegraphics[width=0.7\columnwidth]{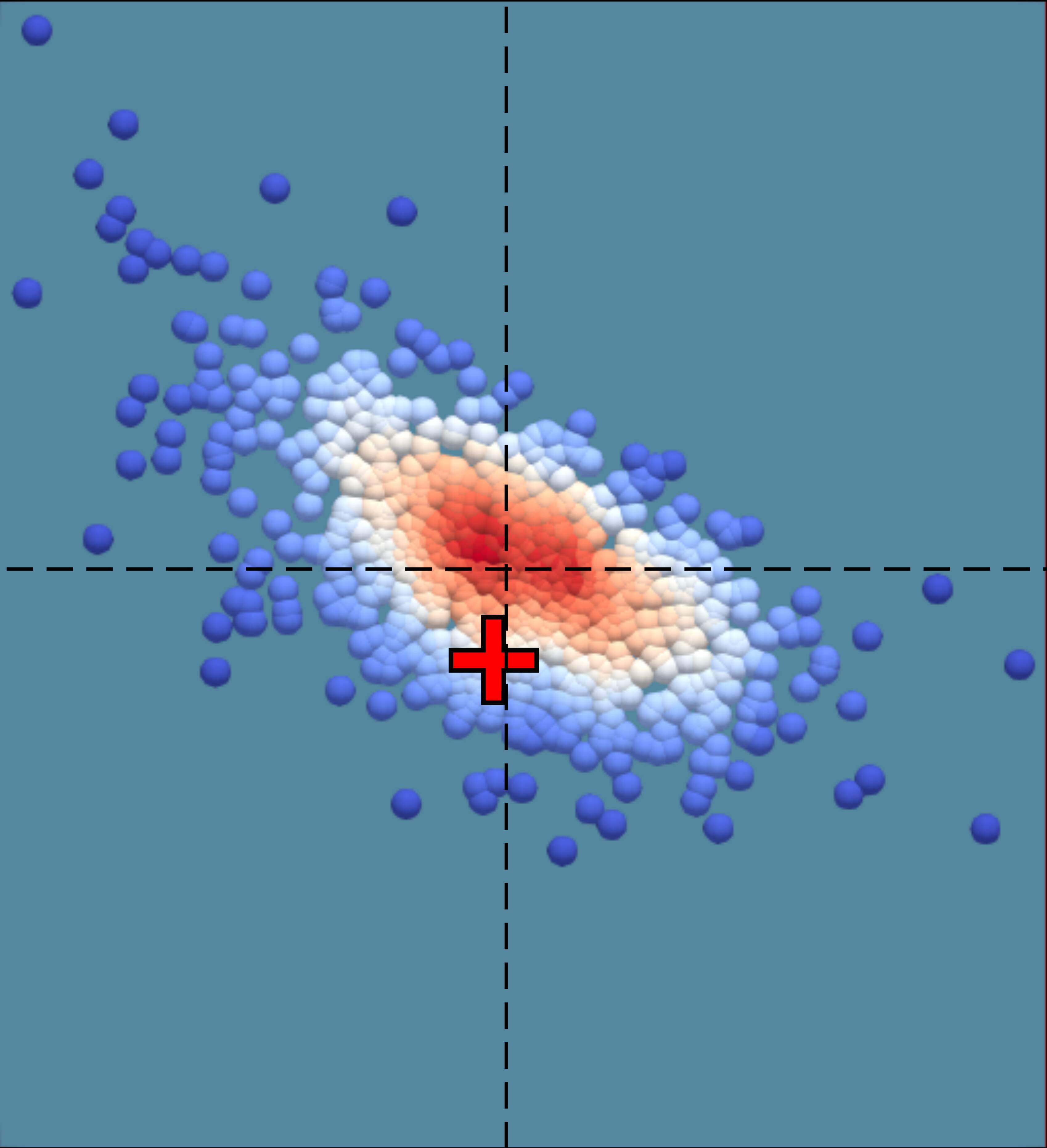}\\
   \includegraphics[width=0.7\columnwidth]{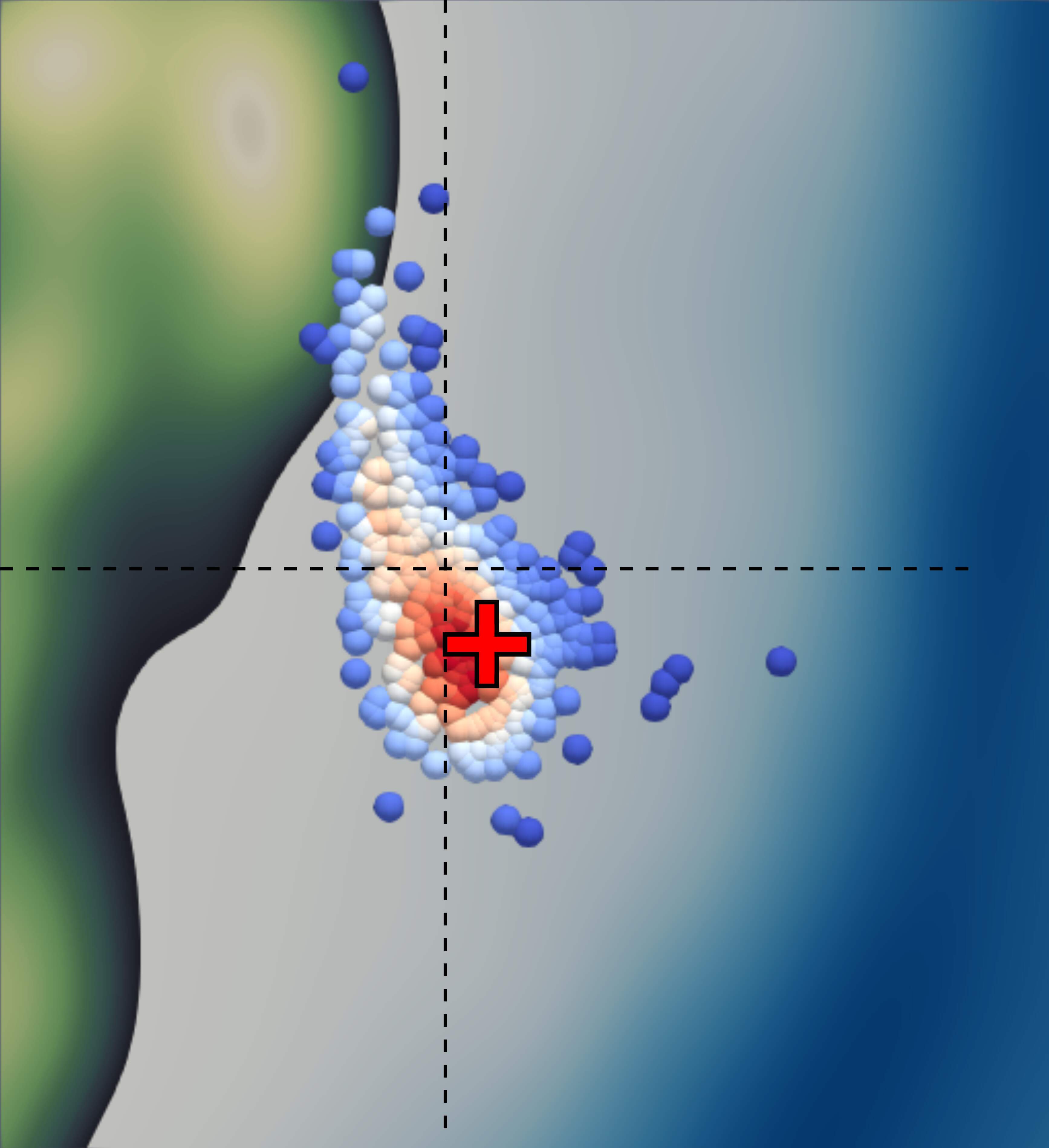}
   \includegraphics[width=0.7\columnwidth]{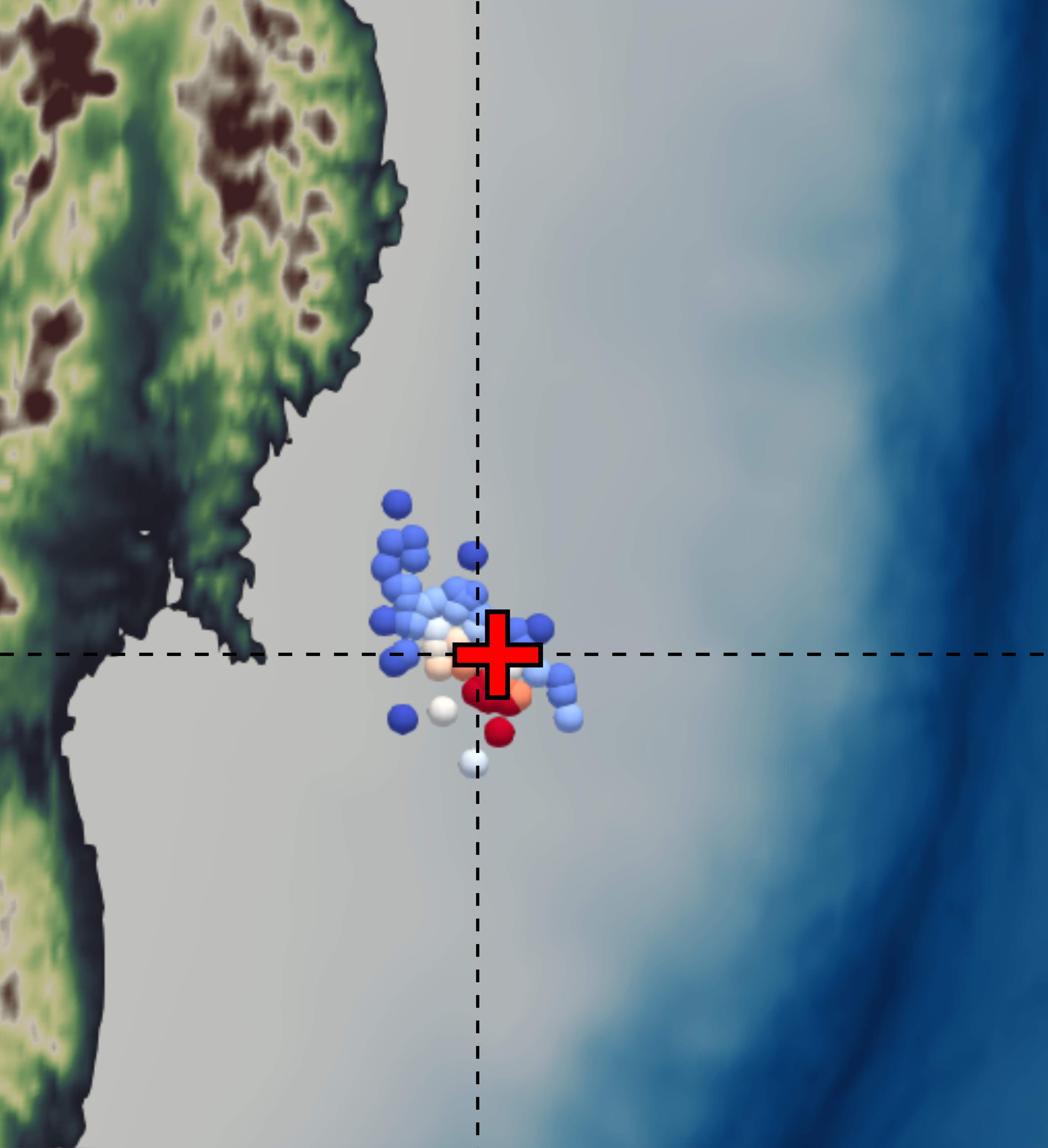}
    \caption{The three-level tsunami test case, each point represents an accepted sample at level $l=0, 1, 2$ The dashed lines show the expected value $\mathbb{E}(Q_0)$ or $\mathbb{E}(Q_0) + \sum_l \mathbb{E}[Q_l - Q_{l-1}]$ with the point $(0,0)$ in red for reference.}
  \label{fig:l0l1}
\end{figure}

%

\begin{figure*}[t]
\centering
   \includegraphics[width=.75\columnwidth]{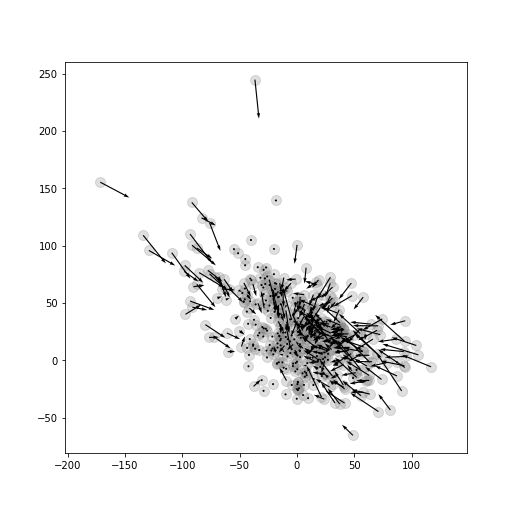}
   \includegraphics[width=.75\columnwidth]{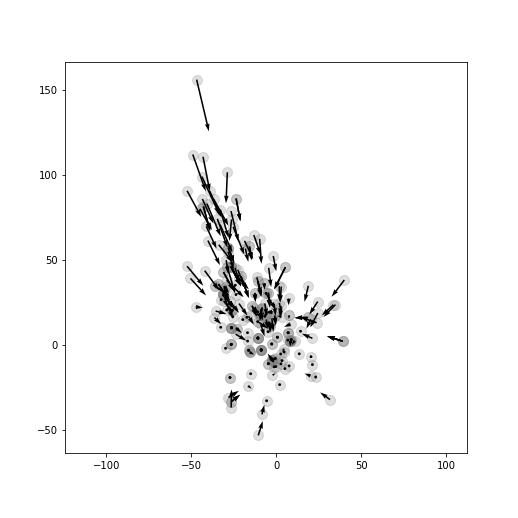}
\caption{\revision{Visualization of samples used to estimate the corrections between levels in \cref{eq:telescoping_sum}.
Left: Correction between levels zero and one. Right: Correction between levels one and two.
Dots and origins of arrows indicate coarse samples. The arrows point towards their corresponding fine samples. Note that the arrow's length is
scaled down to a factor of .15 for clarity. Coarse proposals that were accepted by the fine chain, leading to identical coarse and fine
samples here, appear as simple dots instead of arrows.
}}
\label{fig:corrections}
\end{figure*}

As a second example of the \gls{MLMCMC} method we invert data from the Tohoku tsunami. We find the parameter distribution describing the initial displacements from the data of two available buoys located near the Japanese coast. The scenario is described in more detail in Section \ref{sect:tsunami_model}.
The \gls{MLMCMC} method with three levels computes \revision{$800$ samples on level $0$,  $450$ on level $1$ and $240$ on level $2$ with a subsampling rate of $25$ on level $0$ and $5$ on level $1$. }

These tests were run on up to $72$ Intel Skylake Xeon Platinum 8174 nodes of SuperMUC-ng consisting of $48$ cores each. The tests were run using Intel’s TBB for parallelisation over the $48$ cores of each node and MPI for parallelisation across nodes. Each ExaHyPE run used exactly one full node. \revision{The runtime for each forward model evaluation was on average $7.38$ seconds on level $0$, $97.3$ seconds on level $1$ and $438.1$ seconds on level $2$. These runtimes have a large variablility as the model's timestep depends on the uncertain parameters, making it a challenging test for the scheduling infrastructure. In total $61,250$ level $0$, $34500$ level $1$ and $240$ level $2$ forward model evaluations were required for this test.}

\revision{Again, due to constant parameter dimension across levels,
we only need to choose a proposal density for the
coarsest level. Like before, we choose Adaptive Metropolis \cite{AMMCMC, AMMCMC2} provided by \gls{MUQ}.
As initial prior we set $\mathcal{N}(0, 10 I)$ and update every 100 steps.}

Figure \ref{fig:l0l1} shows the resulting samples on level $0$ and $1$.  
The expected values $\mathbb{E}(Q_0)$, and \revision{$\mathbb{E}(Q_l-Q_{l-1})$} are shown as dashed line. The red marker shows the point $(0,0)$. The point $(0,0)$ is the position of the initial displacements as estimated in \cite{Galvez}.

\revision{In order to illustrate how \gls{MLMCMC} links coarser and finer level posteriors in order to obtain fine level corrections,
\cref{fig:corrections} shows how samples on level $l$ relate to the level $l-1$ samples that served as their respective coarse proposals.
The result can be thought of as a transformation between corresponding coarser and finer distributions, even though only in a non-deterministic sense.
Since in this application we choose the \gls{QOI} to be the uncertain parameter itself, the estimate of the terms in the telescoping sum (\cref{eq:telescoping_sum})
actually correspond to the mean of the corrections displayed here.}

Variances and expected values are given in Table \ref{tbl:ml_tsunami}.  The relatively cheap samples on level $0$ provide a good initial estimate of the posterior, which are improved by the more expensive models utilising the full bathymetric data.
\revision{In contrast to the Poisson case, we do not observe variance reduction across levels.
This is somewhat expected, as the modified bathymetry does not permit the construction of a level hierarchy
fulfilling the theoretical assumptions made by \gls{MLMCMC} based on a priori error estimates. We do,
however, still have the benefit of well-informed proposals on finer levels driven by coarser level chains.}

\begin{table}
\begin{center}
 \begin{tabular}{||c S[table-format=3.2] c S[table-format=4.2] S[table-format=4.2] S[table-format=2.2] S[table-format=2.2]||}
 \hline
                    && &  \multicolumn{2}{c}{$\mathbb{V}[Q_0]$ or} &\multicolumn{2}{c||}{$\mathbb{E}[Q_0] +$}\\
     lvl $l$ &  $t_l$ \revision{[s]} & $\rho_l$ & \multicolumn{2}{c}{$\mathbb{V}[Q_l - Q_{l-1}]$} & \multicolumn{2}{c||}{$\sum_{k=1}^l \mathbb{E}[Q_k - Q_{k-1}]$}\\
 \hline
 \hline
 0 & 7.38 & 25 & 1984.09 &  1337.42 & 3.61 & 27.96 \\
 1 & 97.3 & 5 &1592.17 & 1523.18 & -12.29 &  23.39 \\
 2 & 438.1 & 0 &340.56 & 938.53 & -5.46 & 0.12\\
 \hline
\end{tabular}
\end{center}
\caption{Multilevel properties of the tsunami model. For each level $l$ of mesh, we show the computational cost $t_l$ and chosen subsampling rate $\rho_l$. We show variance and expected values for both components of $Q$.}
\label{tbl:ml_tsunami}
\end{table}

\section{Conclusion}

In this paper we have presented a new parallelization strategy for \gls{MLMCMC} as well as a model-agnostic
implementation as part of the open-source and modular \gls{MUQ} library.
We have demonstrated the effectiveness of algorithm and implementation at solving Bayesian inverse \gls{UQ}
problems on complex and large-scale \gls{PDE} models,
highlighting opportunities in model-specific coarsening strategies in the process.

In order to verify our parallelization approach, we presented both strong and weak scaling results.
An important feature of the work is the applicability of the provided \gls{MUQ} interface, which allows easy coupling to other models and software packages.
This we have shown by applying it to a Poisson model problem implemented in the \gls{DUNE} framework and a tsunami model implemented in the ExaHyPE-Engine.

\begin{acks}
The authors gratefully acknowledge the compute and data resources provided by the Leibniz Supercomputing Centre (www.lrz.de) under project number pr83no. 
We acknowledge funding from the European Union’s Horizon 2020 Programme under the ENERXICO Project, grant agreement No. 828947.

The authors would also like to thank J\"orn Behrens for his extremely useful advice.

We further acknowledge GEBCO Compilation Group (2020) GEBCO 2020 Grid
for providing bathymetry data used in this work (doi:10.5285/ a29c5465-b138-234d-e053-6c86abc040b9).

The authors acknowledge support by the state of Baden-Württemberg through bwHPC
and the German Research Foundation (DFG) through grant INST 35/1134-1 FUGG.
\end{acks}


\bibliographystyle{ACM-Reference-Format}
\bibliography{paper}

\end{document}